\documentclass[preprint]{aastex}
\newcommand{\PSbox}[3]{\mbox{\rule{0in}{#3}\includegraphics{#1}\hspace{#2}}}

\newcommand{\beqa}{\begin{eqnarray}}
\newcommand{\eeqa}{\end{eqnarray}}

\def\gsim {\lower .1ex\hbox{\rlap{\raise .6ex\hbox{\hskip .3ex
        {\ifmmode{\scriptscriptstyle >}\else
                {$\scriptscriptstyle >$}\fi}}}
        \kern -.4ex{\ifmmode{\scriptscriptstyle \sim}\else
                {$\scriptscriptstyle\sim$}\fi}}}
\def\lsim {\lower .1ex\hbox{\rlap{\raise .6ex\hbox{\hskip .3ex
        {\ifmmode{\scriptscriptstyle <}\else
                {$\scriptscriptstyle <$}\fi}}}
        \kern -.4ex{\ifmmode{\scriptscriptstyle \sim}\else
                {$\scriptscriptstyle\sim$}\fi}}}
\def\beq{\begin{equation}}
\def\eeq{\end{equation}}

\def\fig{Figure\ }
\def\figp{Fig.\ }	
\def\figs{Figures\ }
\def\figsp{Figs.\ }
\def\eq{equation\ }
\def\eqp{eq.\ }
\def\eqs{equations\ }

\def\etal{et al.\ }

\def\lya{Ly$\alpha$}

\def\hii{H\,{\sc ii}}
\def\heii{He\,{\sc ii}}
\def\heiii{He\,{\sc iii}}

\def\sec{\, {\rm s}}

\def\kmVs{\, {\rm km\, s^{-1}}}
\def\cm{\, {\rm cm}}

\def\Mpc{\, {\rm Mpc}}
\def\hMpc{\, h^{-1}{\rm Mpc}}
\def\Msun{\, {\rm M_\odot}}

\def\ergVsHz{\, {\rm erg\ s^{-1}\, Hz^{-1}}}

\def\f{\mathtt{f}}
\def\v{\mathtt{v}}

\begin{document}

\slugcomment{{\em Astrophysical Journal, submitted}}
\lefthead{Transverse Proximity Effect}
\righthead{SCHIRBER, MIRALDA-ESCUD\'{E} \& McDONALD}

\title{The Transverse Proximity Effect: A Probe to the Environment,
Anisotropy, and Megayear Variability of QSOs.
}\vspace{3mm}

\author{Michael Schirber${}^1$, Jordi Miralda-Escud\'{e}${}^2$,
 Patrick McDonald${}^3$} 
\affil{${}^1$ Department of Physics, The Ohio State University,
    174 W. 18th Ave, Columbus, OH 43210-1173; 
    schirber@campbell.mps.ohio-state.edu}
\affil{${}^2$ Department of Astronomy, The Ohio State University,
    140 W. 18th Ave, Columbus, OH 43210-1006;
    jordi@astronomy.ohio-state.edu}
\affil{${}^3$ Department of Physics, Jadwin Hall, Princeton University, 
    Princeton, NJ 08544;
    pm@princeton.edu}

\begin{abstract}

The transverse proximity effect is the expected decrease in the
strength of the \lya\ forest absorption in a QSO spectrum when another
QSO lying close to the line of sight enhances the photoionization rate
above that due to the average cosmic ionizing background. We select
three QSOs from the Early Data Release of the Sloan Digital Sky Survey
that have nearby foreground QSOs, with proper line of sight tangential
separations of 0.50, 0.82, and 1.10 $\hMpc$. We estimate that the
ionizing flux from the foreground QSO should increase the
photoionization rate by a factor (94, 13, 13) in these three cases,
which would be clearly detectable in the first QSO and marginally so
in the other two. We do not detect the transverse proximity
effect. Three possible explanations are provided: an increase of the
gas density in the vicinity of QSOs, time variability, and anisotropy
of the QSO emission. We find that the increase of gas density near
QSOs can be important if they are located in the most massive halos
present at high redshift, but is not enough to fully explain the
absence of the transverse proximity effect. Anisotropy requires an
unrealistically small opening angle of the QSO emission. Variability
demands that the luminosity of the QSO with the largest predicted
effect was much lower $10^6$ years ago, whereas the transverse proximity
effect observed in the \heii\ \lya\ absorption in QSO 0302-003 by Jakobsen
\etal (2003) implies a lifetime longer than $10^7$ years. A combination
of all three effects may better explain the lack of Lya absorption
reduction. A larger sample of QSO pairs may be used to diagnose the
environment, anisotropy and lifetime distribution of QSOs.

\end{abstract}

\section{Introduction}

  The \lya\ forest observed in the spectra of QSOs is the principal tool
to study the evolution of the intergalactic medium. Our present
understanding of the \lya\ forest is based on the assumption that the
intergalactic medium is photoionized, and that the absorption features
arise from variations in gas density determined by the gravitational
evolution of primordial fluctuations. Numerical realizations of
Cold Dark Matter models that are supported by independent observational
evidence from the Cosmic Microwave Background, galaxy clustering, and
weak lensing make predictions for properties such as the distribution
of the transmitted flux, the power spectrum, and the transverse sizes
of the absorbers that are in broad agreement with observations (e.g.,
Rauch 1998; Crotts \& Fang 1998; McDonald \etal 2000; Croft \etal 2002;
Tegmark \& Zaldarriaga 2002).

  A fundamental quantity for the \lya\ forest is the intensity of the
ionizing background as a function of redshift, which determines the
degree of ionization of the intergalactic medium. One method for
measuring this intensity is the proximity effect, which consists of a
decline in the strength of the \lya\ absorption as the redshift of the
QSO is approached, compared to the strength observed at the same
redshift in other lines of sight of QSOs with a higher emission
redshift. The effect can be interpreted as due to the additional
ionization caused by the QSO in its vicinity; the total ionizing
radiation is increased near the QSO, reducing the neutral fraction in
the absorption systems and decreasing their column density (Carswell
\etal 1982; Murdoch \etal 1986; Bajtlik, Duncan, \& Ostriker 1988; Scott
\etal 2000). Therefore, the effect can be used to measure the background
intensity: the more intense the background is, the smaller the proximity
effect we expect at a fixed distance from the QSO, because the QSO flux
(known from its luminosity) is less important compared to the mean
background. The most recent measurement by Scott \etal (2000) yields a
value of the photoionization rate of $1.9\times 10^{-12} \sec^{-1}$,
with an error of about 50\%, over the redshift range $1.7 < z < 3.8$.

  This interpretation of the proximity effect to measure the ionizing
background intensity is, however, subject to several systematic
uncertainties: (a) Systematic errors in the redshift of the QSO
estimated from its emission lines may change the distances inferred
from the absorption systems to the QSO. Scott \etal show that the
redshifts inferred from the O\,{\sc iii} and Mg\,{\sc ii} emission
lines are higher by $400 \kmVs$ than the redshift obtained from the \lya\
emission line, resulting in a factor of 2 decrease in the inferred
background intensity. (b) QSOs could be variable on the
photoionization timescale of $\sim 10^4$ years, and the proximity
effect reflects the QSO luminosity averaged over a photoionization
timescale, which would be systematically lower than the present
luminosity because QSOs in a flux-limited sample should be
preferentially selected in their bright phase. (c) Some QSOs may be
gravitationally lensed, again making them appear on average more
luminous than they really are in a flux-limited sample. (d) QSOs are
likely associated with dense regions in the universe, and the higher gas
density in their vicinity may partially compensate for the decrease in
the neutral fraction due to their ionizing flux.

  In fact, in the context of the \lya\ forest theory based on primordial
fluctuations, the value of the ionizing background intensity can be
determined from the observed transmitted flux distribution, once the
mean baryon density, gas temperature and power spectrum amplitude are
known, because numerical simulations allow one to predict the density
and temperature distribution of the gas in the \lya\ forest. This method
yields a value of the ionizing background intensity that is lower than
that found from the proximity effect by a factor of 2 to 3 (e.g.,
McDonald \& Miralda-Escud\'e 2001; Schirber \& Bullock 2003).

 Several authors have searched for the transverse proximity effect in
the past, starting with Crotts (1989), who observed a triplet of QSOs
at $z\sim 2.5$ with separations of 2 to 3 arc minutes, corresponding
to time-delays of a ${\rm few}\times10^6$ years.  No convincing
evidence was found for the proximity effect in this and later
observations (M\o ller \& Kj\ae rgaard 1992; Fern\'andez-Soto \etal
1995; Liske \& Williger 2001). One exception is the case of Q0302-003,
where a void of absorption lines was found by Dobrzycki \& Bechtold
(1991) in the vicinity of a QSO; however, the center of the void is
offset by $\Delta z\approx0.05$ to the blue of the QSO, which
corresponds to a surprisingly large separation of $\sim8h^{-1}$ proper
Mpc. These analyses, with detections and otherwise, have been based on
taking some measure of the absorption in the region where the proximity
effect is expected, and comparing this to an empirical model of
absorption systems derived from other QSO line of sights.  Most often
the absorption has been characterized by the number of \lya\ lines with
fitted Voigt profiles that are found per unit redshift. A different
approach was used by M\o ller \& Kj\ae rgaard and Liske \& Williger:
they measured the mean transmitted flux distribution in the proximity
effect region and compared it to a modeled transmitted flux
distribution, derived by assuming that the \lya\ forest consists of
uncorrelated absorption lines with a distribution of column densities
and Doppler parameters obtained from fits to the observed spectra. In
this paper, we use the full transmitted flux distribution as a probe to
the transverse proximity effect, comparing it instead to the expected
distribution obtained from artificial absorption spectra generated from
cosmological simulations of the \lya\ forest.

  This paper presents an analysis of three QSO pairs found in the
Sloan Digital Sky Survey Early Data Release, which are close enough
for the expected transverse proximity effect to be important. The
sample selection and continuum fitting of the spectra is described in
\S 2. The method we use to estimate the probability of observing a
certain value of the transmitted flux in the spectrum, with the use of
\lya\ forest simulations, is described in \S 3. By comparing the
actual spectra to simulated lines of sight, we show in \S 4 that there
is no evidence for the transverse proximity effect in our three
pairs. We then model the possible effect of a density enhancement in
the vicinity of the QSOs (\S 5), and discuss the effects of anisotropy
(\S 6) and QSO variability (\S 7). We assume the flat cosmological
model with present matter density $\Omega_m = 0.4$ and Hubble constant
$H_0 = 65 \kmVs\Mpc^{-1}$ throughout the paper.

\section{Sample}

  The SDSS Early Data Release contains a total of 3814 QSOs with
redshifts between 0.15 and 5.03, and magnitudes $15.16 < i^* < 20.82$
(Schneider \etal 2002). We first took from these the 482 QSOs with
$z>2.2$, on which the \lya\ forest spectrum is (at least partly) in the
observed wavelength range. For each QSO we searched for the closest
neighbor in angular separation, and selected those pairs for which the
foreground and background QSO redshifts, $z_f$ and $z_b$, satisfied
$1+z_f > (1026/1216)\cdot(1+z_b)$, to ensure that the proximity effect
region would not be contaminated by Ly$\beta$ lines.

  Of the pairs that fulfilled this criterion, we selected the three
with the largest ratio of the ionizing rate from the foreground
QSO, $\Gamma_f$, to that of the background ionizing rate,
$\Gamma_{bkg}$, evaluated on the line of sight of the background QSO
at a redshift $z_f$. The calculation of these ionizing rates is
described below (\S 2.1). We list these 3 pairs in Table 1, along with
their redshifts and angular separations, from which we obtained their
impact parameter (i.e., the shortest distance between the foreground
QSO and the background QSO line of sight). For this distance, we also
determined the expected continuum flux from the foreground QSO at the
Lyman limit (see \S\ref{sect:cont_fit}).  

\begin{table*}[ttt] 
\begin{center}
\begin{tabular}{cccccccc}
   \tableline         
Pair                             &
Background QSO                   & Foreground QSO             &
 $z_b$             & $z_f$       & $\theta$     & $R_\bot$    &
 $\f_\bot$
\\ \tableline\tableline
1                                &
$110819.15-005823.9$             & $110813.85-005944.5$       &
 4.595             & 4.032       & 1.89         & 0.50        &
 14.3                            
\\ 
2                                &
$134755.67+003935.0$             & $134808.79+003723.2$       &
 3.817             & 3.620       & 3.95         & 1.10        &
 2.29 
\\ 
3                                &
$233132.84+010620.8$             & $233139.76+010427.0$       &
 2.639             & 2.245       & 2.57         & 0.82        &
 4.48 
\\ \tableline
\end{tabular}
\end{center}
Table 1 -- The pairs in our sample. {\it Columns 2 and 3:} Sloan names
for background and foreground QSOs. {\it Columns 4 and 5:} redshifts
for both QSOs. {\it Column 6:} the angular separation between the
pairs in arc minutes. {\it Column 7:} impact parameter in (proper)
$h^{-1}{\rm Mpc}$. {\it Column 8:} the Lyman limit flux density in
units of $4\pi\times10^{-21} {\rm erg/s/cm^2/Hz}$ at a distance of
$R_\bot$ from the foreground QSO.  In terms of the notation in
\S\ref{sect:Gamma_Q} , this is equal to $\f_f(\nu_{\rm H},z_f)$. Note that
$\f_\bot=1$ corresponds to an ionizing background intensity of
$J=10^{-21} \rm{erg/s/cm^2/Hz/sr}$.
\end{table*} 

\subsection{Continuum fitting}
\label{sect:cont_fit}

\begin{figure}[th!]
  \PSbox{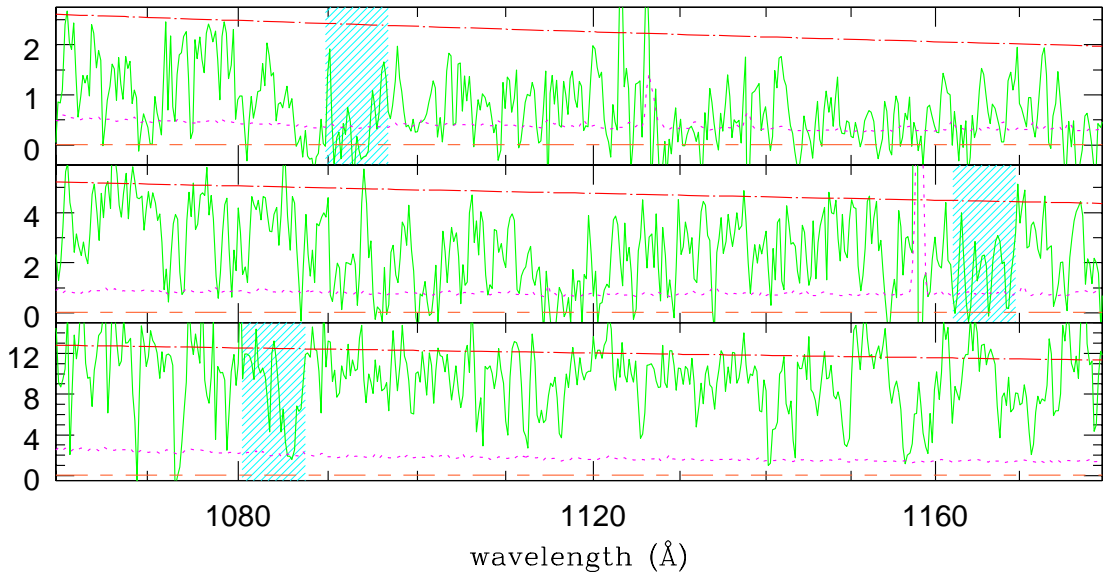  hoffset=-20 voffset=-380 hscale=85 vscale = 85}
  {150mm}{75mm}  
\caption{{\it Solid lines:} spectra of background QSOs (in units of
$10^{-17}\,{\rm erg/s/cm^2/\mbox{\AA}}$) versus rest-frame wavelength.
\emph{Top} panel is pair 1, \emph{middle} is pair 2, \emph{bottom} is
pair 3. {\it Dot-dash line:} continuum fit; {\it dotted line:} noise.
Shaded regions are centered at the foreground QSOs redshifts, with
rest-frame widths $\Delta \v=2000\,{\rm km/s}$.
\label{fig:spec_comp} }
\end{figure}

  To measure the fraction of transmitted flux from the observed QSO
spectral flux, one needs to first estimate the continuum flux that would
have been observed in the absence of any \lya\ absorption. Here, we use a
method similar to the one proposed by Press, Rybicki, \& Schneider
(1993), based on extrapolating the flux observed at wavelengths longer
than \lya\ to the shorter wavelengths where the \lya\ forest absorption
is present. The other method that is often used is to assume that the
spectral regions between absorption lines are practically free of
absorption and can be used to trace the continuum (e.g., Rauch \etal
1997), but this becomes more problematic at high redshift where
absorption becomes ubiquitous.

  We fitted the continuum of each one of the six QSO spectra in the
three pairs, using only the wavelength range from the rest-frame \lya\ to
the observer frame wavelength limit of 8500\AA\, and eliminating the
wavelength intervals corresponding to the strong emission lines
(Ly$\alpha$, O\,{\sc i}, Si\,{\sc iv}-O\,{\sc iv}, C\,{\sc iv},
C\,{\sc iii}, Fe\,{\sc iii}, Fe\,{\sc ii}) that are given in Table 2 of
Vanden Berk \etal (2001). The remaining part of each
spectrum was fit with a single power-law, $\f_\lambda\propto
\lambda^{-\alpha_\lambda}$. We provide the spectral slope and the
fitted continuum flux per unit wavelength at rest-frame 1460 \AA\
in Table 2 for the background QSOs, and in Table 3 for the
foreground QSOs. We plot our fits to the background spectra,
extrapolated to wavelengths shorter than \lya, in \fig
\ref{fig:spec_comp}. The observed flux is
divided by our continuum to obtain the fraction of transmitted flux,
$F$. The average transmitted flux, $\bar{F}$,
over the \lya\ forest region covering
the QSO rest-frame wavelength range from 1060 \AA\ to 1180 \AA\ (to
avoid the emission lines at \lya\ and Ly$\beta$) is given in Table 2.
We also give the mean transmitted flux measured by Bernardi \etal
(2003) over this same region. In terms of
$\tau_{eff}\equiv-\ln(\bar{F})$, they measured:\footnote{Bernardi
\etal (2003) also include a small dip in the effective optical depth
at $z\sim3$ that they attribute to helium reionization. We have not
included this in our analysis.}

\beq
  \tau_{eff} = 2.4\times10^{-3}(1+z)^{3.79\pm0.18}
\label{eq:tau_eff_B}
\eeq 

\noindent
Our values of the average transmission are similar to those of
Bernardi \etal (2003). We will also use the Bernardi \etal (2003)
result later in \S 3 to normalize the optical depth in the numerical
simulations of the \lya\ forest.

\begin{table*}[ttt]
\begin{center}
\begin{tabular}{cccccc}
   \tableline          
Pair                             & Background spectrum        &
 $\f_b(1460,0)$                  & $\alpha_\lambda$           &
 $\bar{F}_{obs}$                 & $\bar{F}_{exp}$
\\ \tableline\tableline
1                                & spSpec-51900-0278-215      &
 1.13                            & $2.62\pm0.33$              &
 0.31                            & 0.30                         
\\ 
2                                & spSpec-51666-0300-414      &
 3.06                            & $1.66\pm0.07$              &
 0.53                            & 0.51                       
\\ 
3                                & spSpec-51821-0384-334      &
 8.90                            & $1.13\pm0.02$              &
 0.80                            & 0.79                       
\\ \tableline

\end{tabular}
\end{center}
Table 2 -- {\it Column 2:} background QSO spectrum name. {\it Column 3:}
flux density at rest-frame 1460 \AA\ in
$10^{-17}{\rm erg/s/cm^2/\mbox{\AA}}$ for our continuum fit (errors
$\le 2\%$). {\it Column 4:} spectral slope (with error). {\it Column 5:}
transmitted flux averaged over $1060-1180\,\mbox{\AA}$. {\it Column 6:}
mean transmitted flux from Bernardi \etal (2003) averaged over the same
wavelengths (see \eqp \ref{eq:tau_eff_B}).  
\end{table*}

\subsection{The Ionizing Flux from the Foreground QSO}
\label{sect:Gamma_Q}

  The signature of the proximity effect is an increase in transmitted
flux in the background QSO spectrum. Photoionization equilibrium of the
hydrogen in the IGM (neglecting collisional ionization) implies that the
optical depth is inversely proportional to the photoionization rate:
$\tau\propto\Gamma^{-1}$. The addition of flux from the foreground QSO
decreases the optical depth by a factor $(1+\omega)$, where $\omega$ is
the ratio of the ionization rate from the foreground QSO, $\Gamma_f$, to
the diffuse background ionization rate, $\Gamma_{bkg}$. Here, we give
the equations to calculate $\Gamma_f$, and in the following subsection
we describe the background ionization rate that we employ.

We assume for now that the luminosity, $L_f(\nu)$, of the foreground
QSO is constant and isotropic.  The flux along the line of sight to
the background QSO is then $\f_f(\nu,z) = L_f(\nu)/[4\pi R_f^2(z)]$,
where $\nu$ is the rest-frame frequency, and $R_f(z)$ is the proper
distance between the foreground QSO and the point where the flux is
measured given by:

\beq 
  R_f = \sqrt{R_\bot^2+R_\parallel^2} ~,
\label{eq:R_f}
\eeq

\noindent
where $R_\bot$ is the impact parameter from Table 1, and

\beq
  R_\parallel \equiv \frac{c}{H(z_f)} \ \frac{z-z_f}{1+z_f} 
              \equiv \frac{\Delta \v}{H(z_f)} ~.
\label{eq:R_ll}
\eeq 

\noindent
Because $R_f$ is small, we have neglected the redshift correction to
the flux $\f_f(\nu,z)$.  The luminosity of the foreground QSO
is $L_f(\nu)=4\pi d_L^2(z_f) \ \f_f(\nu_0,0)$, where
$d_L(z_f)$ is the luminosity distance, and $\f_f(\nu_0,0)$ is the
observed flux at $\nu_0=\nu/(1+z_f)$.  We extrapolate our continuum
fit to obtain the observed flux at the rest-frame Lyman limit,
$\nu_{\rm H}=13.6{\rm eV}$.  This value can be found in Table 3, along
with the Lyman limit luminosity, for each foreground QSO.  In
addition, the Lyman limit flux at closest separation, $\f_f(\nu_{\rm
H},z_f)$, is written in Table 1.  The ionizing flux beyond the Lyman
limit is related to the photoionization rate by: $\Gamma_f(z) =
\int^\infty_{\nu_{\rm{H}}} \f_f(\nu,z)\, \sigma_{\rm H}(\nu)/(h_p\nu)
\,d\nu$, where $\sigma_{\rm H}(\nu)$ is the hydrogen photoionization
cross section and $h_p$ is the Planck's constant. Assuming that the QSO
spectrum shortward of the Lyman limit can be characterized by a single
power-law, $L_f(\nu)\propto\nu^{-1.57}$ (Telfer \etal 2001, for their
radio-quiet sub-sample), then

\beq 
  \Gamma_f(z) = 0.21\times 10^{-12}\,
    {\rm s}^{-1}\, \left(\frac{\f_f(\nu_{\rm H},z)}{10^{-21}\, 
    {\rm ergs/s/cm^2/Hz}}
  \right) ~.
\label{Gammafz}
\eeq 

\begin{figure}[th!]
  \PSbox{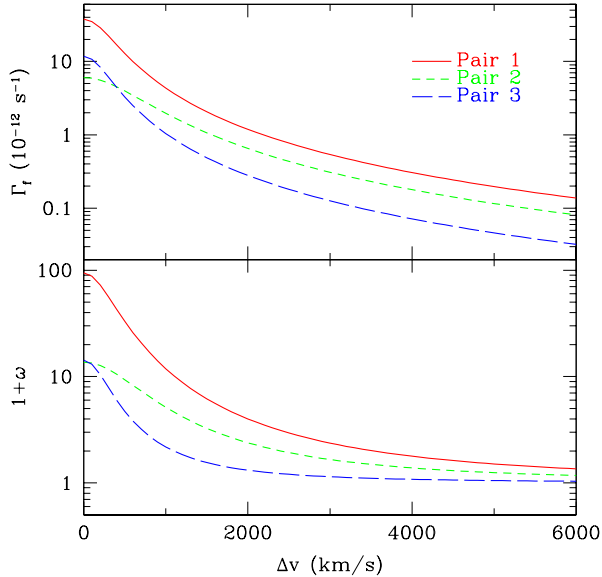  hoffset=70 voffset=-65 hscale=40 vscale = 40}
  {75mm}{75mm}  
\caption{ \emph{Top}: Photoionization rate from the foreground QSOs
in our sample versus velocity separation,
$\Delta \v\equiv c(z-z_f)/(1+z_f)$ (the curves are essentially
symmetric about $\Delta \v = 0$).
\emph{Bottom}: Ratio of photoionization rate contributed by the
foreground QSO to that of the background QSO,
$\omega=\Gamma_f(z)/\Gamma_{bkg}$.
\label{fig:G_n_w} }
\end{figure}

\noindent 
The photoionization rates for each of our QSO pairs are plotted in
\fig \ref{fig:G_n_w}, as a function of the velocity separation between
$z_f$ and $z$. Also plotted are the ratios,
$\omega\equiv\Gamma_f/\Gamma_{bkg}$, assuming the background rate in
\S\ref{sect:Sim}. The maximum $\omega$ values, at the redshift of the
foreground QSO, for each pair can be found in Table 3. The largest
proximity effect is expected in pair 1 where, because of the small
angular separation, the high luminosity of the foreground QSO, and the
relatively low background intensity, $\omega$ is expected to be as
large as $\sim 90$ and to be substantially above one up to redshift
separations of a few thousand km/s.

\begin{table*}[ttt]
\begin{center}
\begin{tabular}{cccccccc}
   \tableline          
Pair                             & Foreground Spectrum        &
 $i^*$                           & 
 $\f_f(1460,0)$                  & $a_\lambda$                &
 $\f_f(\nu_{\rm H},0)$           & 
 $L_f(\nu_{\rm H})$              &
 $\omega(z_f)$
\\ \tableline\tableline
1                                & spSpec-51900-0278-203      &
 19.28                           &
 2.75                            & $1.67\pm0.07$              &
 4.23                            &    
 5.41                            &
 93.7                            
\\ 
2                                & spSpec-51666-0300-461      &
 19.34                           & 
 3.42                            & $1.36\pm0.05$              &
 3.84                            &    
 4.14                            &
 12.6
\\
3                                & spSpec-51821-0384-338      &
 18.46                           &
 13.0                            & $1.87\pm0.01$              &
 9.18                            &
 4.56                            &
 13.3
\\ \tableline
\end{tabular}
\end{center}
Table 3.  {\it Column 2:} foreground QSO name. {\it Column 3:} $i^*$
magnitude. {\it Column 4:} flux at rest-frame 1460 \AA\ in
$10^{-17}{\rm erg/s/cm^2/\mbox{\AA}}$ for our continuum fit
(errors $\le 2\%$). {\it Column 5:} spectral slope (with error).
{\it Column 6:} flux at rest-frame Lyman limit in
$10^{-28}{\rm erg/s/cm^2/Hz}$, using our continuum fit. {\it Column 7:}
luminosity at the Lyman limit in $10^{30} h^{-2} \ergVsHz$.
{\it Column 8:} ratio of photoionization rates, 
$\omega=\Gamma_f/\Gamma_{bkg}$, at the redshift of the foreground QSO.
\end{table*}

\section{Simulation}
\label{sect:Sim}

  This section describes the simulations that we use to evaluate the
probability of observing a certain value of the transmitted flux, for
an assumed photoionization rate enhanced by a foreground QSO
calculated as described in the previous section.

\begin{figure}[th!]
  \PSbox{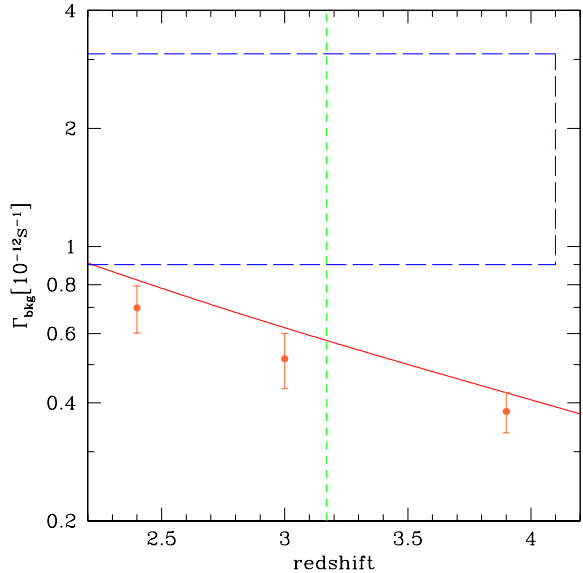  hoffset=70 voffset=-65 hscale=40 vscale = 40}
  {75mm}{75mm}   
\caption{The solid line is our adopted background
ionization rate.  Note that in generating this background we have
extrapolated the Bernardi \etal (2003) fit to $\tau_{eff}$ slightly
beyond the data limits ($z=2.6-4.0$). The vertical line at $z=3.17$
shows the redshift of the simulation output we analyze. The dashed box
shows the limits from Scott \etal (2000) using the line-of-sight
proximity effect.  The data points are from McDonald \&
Miralda-Escud\'e (2001, 2003); their lower rate compared to the one
adopted here is due to their higher temperature and power spectrum
amplitude, as discussed in the text.  
\label{fig:Gamma_eff} }
\end{figure}

  We use an HPM simulation with $512^3$ cells and a comoving box length
of $40h^{-1}{\rm Mpc}$ (for a description of HPM simulations see Gnedin
\& Hui 1998). From the resulting gas density, $\rho_{gas}$, the optical
depth is calculated assuming photoionization equilibrium and using the
temperature-density relation $T = 1.1\times10^4 \, (\rho_{gas} /
\bar\rho_{gas})^{0.3} \, {\rm K}$. For the average gas density,
$\bar\rho_{gas}$, we assume $\Omega_b h^2 = 0.02$ (O'Meara \etal 2001).
The one free parameter is $\Gamma_{bkg}(z)$, which we set so that the
average transmitted flux, $\bar{F}$, is in agreement with the results
from Bernardi \etal in \eq (\ref{eq:tau_eff_B}). Our effective
background rate is plotted in \fig \ref{fig:Gamma_eff}.

  The density distribution in the simulation is generated from a flat
CDM model with $\Omega_m=0.4$, $h=0.65$, $\sigma_8=0.83$ (consistent
with cluster abundances), and primordial spectral index $n_s=0.85$. In
comparison to the power spectrum determination from the \lya\ forest in
Croft \etal (2002), who found an amplitude of $\Delta^2(k_p) =
0.74^{+0.20}_{-0.16}$ at $z=2.72$ for $k_p=0.03(\kmVs)^{-1}$, our power
spectrum has $\Delta^2(k_p)=0.62$ [where $\Delta^2(k)=k^3P(k)/(2\pi^2)$,
and $P(k)$ is the power spectrum]. Thus, our power spectrum amplitude is
on the low side of present estimates (note that Seljak, McDonald, \&
Makarov 2003 find that the power spectrum amplitude may be higher than
found by Croft et al.~). A lower amplitude of the power spectrum implies
that there is more gas left in voids, hence providing more absorption
over most regions in the \lya\ spectrum. This means that to obtain the
same mean transmitted flux, we require a \emph{higher} value of the
background photoionization rate, $\Gamma_{bkg}$. Moreover, the gas
temperature we assume is on the low side of the observational
determinations (Theuns \etal 2002 and references therein), and a lower
temperature also leads to a higher photoionization rate because of the
increased recombination coefficient. Hence our value of $\Gamma_{bkg}$
is actually an upper limit inferred from the mean transmitted flux,
giving a lower limit for $\omega = \Gamma_f/\Gamma_{bkg}$, and therefore
making our predicted proximity effect conservatively low. 

  We also show in \fig \ref{fig:Gamma_eff} the measurement of
$\Gamma_{bkg}$ by Scott \etal (2000) from the line-of-sight proximity
effect. This value is higher than our estimate by a factor of at least
$\sim 2$ at high redshift. However, as mentioned in the introduction
there are several possible systematic errors in this measurement, in
particular the effect of the gas overdensity near the QSOs; see \S 5
below.

\subsection{Resolution, Pixel Size, and Noise}

  Once we have chosen the background, we can generate simulated spectra
with and without the effect of the foreground QSO. But before comparing
these to the Sloan spectra, we have to match the resolution, pixel size,
and signal-to-noise of the data.   

  We first smooth each line of sight in the simulation separately with
the average resolution quoted in Schneider \etal (2002):
$\lambda/\Delta\lambda=1950$. We then adjust the pixel size in the
simulation to match the pixel size of the Sloan spectra, which is
$d\v_{spec}=c\cdot d \lambda_o/\lambda_o =69{\rm km/s}$. The
pixel size of the simulation corresponds to a velocity range of
$4.9 {\rm km/s} \ (1+z)^{1/2}$, so we simply average the flux over
$14 \ (1+z)^{-1/2}$ pixels. Finally, we add noise to the simulated
spectra by adding to the transmitted flux a random number selected from
a Gaussian distribution centered at zero with width equal to
$N(\lambda)/\f_b(\lambda)$,
where $N(\lambda)$ is the noise supplied for each Sloan spectrum, and
$\f_b(\lambda)$ is our continuum fit.  

\section{Results}

\begin{figure}[th!]
  \PSbox{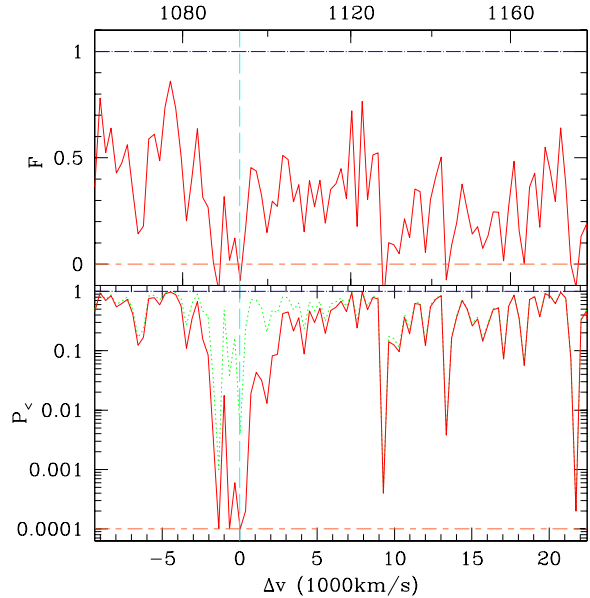  hoffset=70 voffset=-65 hscale=40 vscale = 40}
   {75mm}{75mm} 
\caption{ \emph{Top}: Observed transmitted flux in the background QSO
of pair 1, versus velocity separation from the foreground QSO, after
rebinning the spectrum in \fig \ref{fig:spec_comp}  over 5
pixels. \emph{Bottom}: Fraction of 5000 simulated spectra with
transmitted flux below the measured value. The dotted line assumes a
uniform ionization rate, and the solid line includes the ionizing flux
from the foreground QSO.  If \emph{none} of our 5000 simulated spectra
have transmitted flux below the observed one, we plot the
corresponding point along the dashed line at $10^{-4}$.
\label{fig:FnP_215} }
\end{figure}

\begin{figure}[th!]
  \PSbox{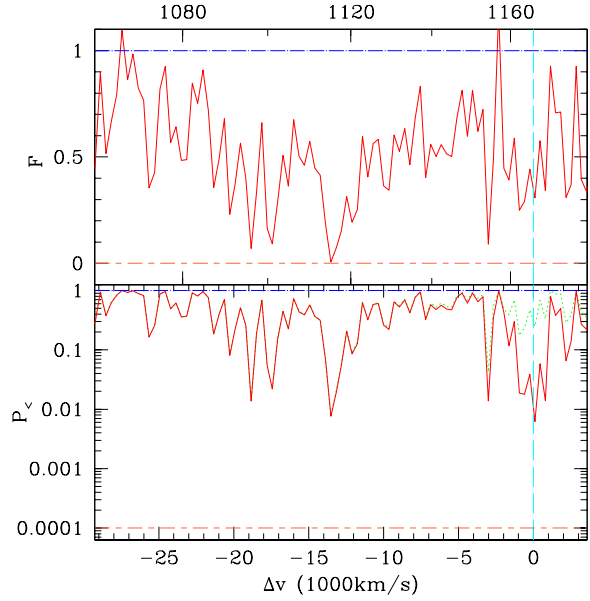  hoffset=70 voffset=-65 hscale=40 vscale = 40}
  {75mm}{75mm}  
\caption{
Same as \fig \ref{fig:FnP_215} for pair 2.  
\label{fig:FnP_414} }
\end{figure}

\begin{figure}[th!]
  \PSbox{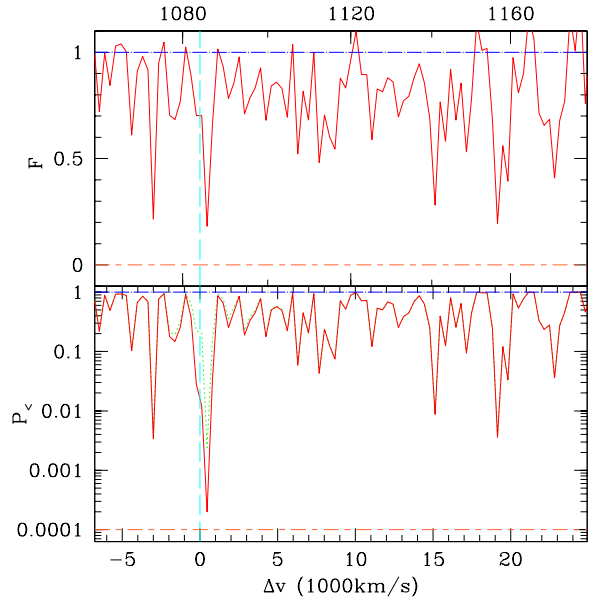  hoffset=70 voffset=-65 hscale=40 vscale = 40}
  {75mm}{75mm}
\caption{ 
Same as \fig \ref{fig:FnP_215} for pair 3. 
\label{fig:FnP_334} }
\end{figure}

  Once we have smoothed, rebinned, and added noise to our simulated
spectra, we are ready to compare them to the real spectra. We make
this comparison with two sets of simulated spectra:  one with the
background ionization rate only (``Quasar Off''), and the other with
the additional ionization rate due to the foreground QSO (``Quasar
On''). To test the presence of the transverse proximity effect, we
plot in \figs \ref{fig:FnP_215}, \ref{fig:FnP_414}, and
\ref{fig:FnP_334}, for each one of our three QSO pairs, the fraction
$P_<$ of simulated spectra in which the transmitted flux is below the
measured value in the observed spectrum. If the modeling is correct,
then half of the simulated spectra should have a transmitted flux
below the measured one.  In these plots, we have averaged the observed
spectra over 5 pixels, corresponding to $\Delta \v=345{\rm
km/s}$. At this velocity separation, the \lya\ forest correlation
function is small and so the transmitted flux values in adjacent
pixels are close to independent.

  Far from the redshift of the foreground QSOs, the probabilities
$P_<$ are in fact not far from uniformly distributed. But in the
vicinity of the foreground QSOs, the fraction of ``Quasar On'' spectra
below the observed spectrum is very small for all three pairs. In
other words, the evidence shows that the transverse proximity effect
is absent in all three cases.  There is no large reduction in
absorption observed at the redshift of the foreground QSO as would be
expected from the increased ionizing flux predicted in \fig
\ref{fig:G_n_w}.  Indeed, there is actually strong \lya\ absorption in
the vicinity of the foreground QSO in pairs 1 and 3, and about average
transmission in pair 2.

One concern is that the measured redshifts of the foreground QSOs
may suffer from the same systematic uncertainties that plague the line-of-sight
proximity effect.  However, the SDSS redshifts do not depend on a single
emission line but instead are determined by
cross-correlating the continuum-subtracted spectra (see Stoughton \etal 2002)
with the composite spectrum from Vanden Berk \etal (2001).
For reference, the foreground QSO reshifts of pair 1, 2, 3
have, respectively, errors of
$39\kmVs$, $54\kmVs$, $136\kmVs$.
Since these errors correspond to less than two pixels in the spectra, we
will neglect them in the rest of the analysis.
Even if these errors were underestimated, our visual inspection of the
spectra suggests that the errors cannot be large enough to change our
conclusions on the transverse proximity effect, which should be detectable
over several thousands of km/s.

\subsection{Discrete Absorbers}

  Before we proceed to discuss the possible explanations for the
absence of the proximity effect, we must address a caveat with our
estimate of the probability, $P_<$, that the transmitted flux is below
the observed value, based on simulations of the \lya\ forest spectra.
In the presence of the additional ionizing flux from the foreground
QSO, any optically thick absorption line will require an absorber with a
high enough column density such that, even after dividing it by the
factor $1+\omega$ (in \figp \ref{fig:G_n_w}), the absorber still produces
the low observed transmission. For example, to get the saturated
absorption we see in \fig \ref{fig:FnP_215} for pair 1, where we have
averaged over 5 pixels, would require a column density of $\sim
10^{15} \cm^{-2}$ (assuming a minimum \lya\ optical depth of 3 over
the velocity width: $\Delta \v=345\kmVs$).  But if the
ionizing flux has been increased by a factor $1+\omega$ due to the
proximity of a QSO, the absorber must have had a minimum column
density $10^{15} (1+\omega) \cm^{-2}$ before the QSO turned on.

  However, we know that the abundance of absorbers with this high column
density is underestimated by a factor of 3 to 10 in simulations similar
to the one used here, because of the limited spatial and mass resolution
(Miralda-Escud\'e \etal 1996; see also Gardner \etal 1997, 2001 for a
detailed study of Lyman limit systems in simulations of different models
and the effect of resolution). The probability of finding strong
absorption in the simulations may therefore be too low, particularly
when the flux of the foreground QSO is taken into account. Nevertheless,
the probability $P_<$ in the pixels near the foreground QSO redshift is
sufficiently small, especially in pair 1, that correcting for this would
not change the high improbability that a sufficiently strong
absorber just happens to be in the vicinity of the foreground QSO. 

\begin{figure}[th!]
  \PSbox{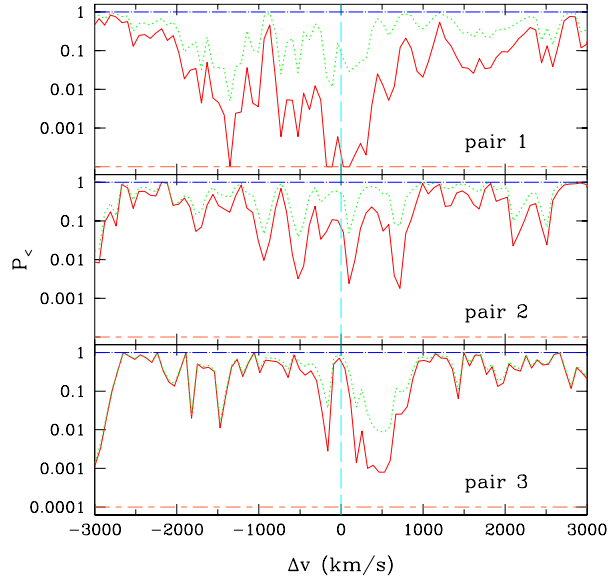  hoffset=70 voffset=-65 hscale=40 vscale =
   40} {75mm}{75mm} \caption{ Same as bottom panels in \figs
\ref{fig:FnP_215}, \ref{fig:FnP_414}, \ref{fig:FnP_334}, but for single
pixels in the SDSS spectra. {\it Solid lines}: ``Quasar On''
simulations. {\it Dotted lines}: ``Quasar Off''. 
\label{fig:prob_tran_comp} }
\end{figure}

\begin{figure}[th!]
  \PSbox{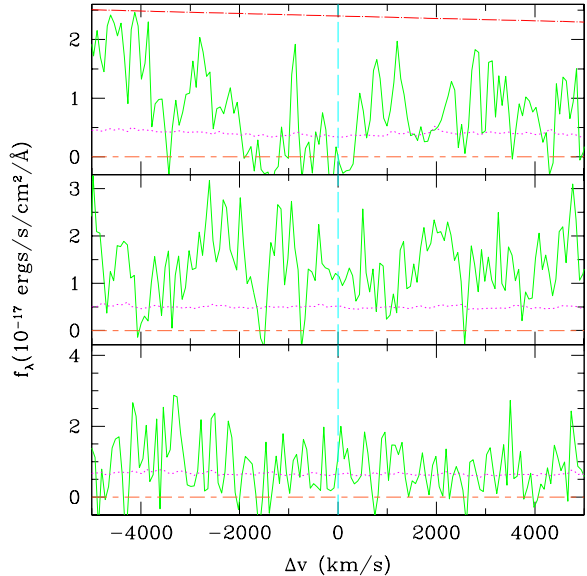  hoffset=70 voffset=-65 hscale=40 vscale = 40}
   {75mm}{75mm} \caption{ Spectrum of the background QSO in pair 1
   (solid lines) centered at the redshift of the foreground QSO
   (vertical dashed lines) of \lya\ (top panel), Ly$\beta$ (middle
   panel), and Lyman limit (bottom panel). The dotted lines are the
   noise, and the dot-dashed line is the continuum fit. Note the absence of
   Lyman limit absorption or strong Ly$\beta$ absorption near the
   foreground QSO redshift.
\label{fig:a_v_bl_215} }
\end{figure}

  Furthermore, the high column density absorber required to produce the
\lya\ absorption at the redshift of the foreground QSO in the case of
pair 1 would also need to have an unusually high velocity dispersion.
In \fig \ref{fig:prob_tran_comp}, $P_<$ is replotted for the three pairs
with the original pixel size (i.e., without averaging over 5 pixels).
For pair 1, the low probability of the observed \lya\ flux after the QSO
ionization is taken into account (solid line) is spread over a wide
velocity range of $\sim 2000 \kmVs$. This very large velocity spread is
rare among high column density systems (Prochaska \& Wolfe 1997), and
when very high velocity dispersions occur (e.g., Bahcall \etal 1996),
the hydrogen column density tends to be concentrated on narrow
subcomponents. \fig \ref{fig:a_v_bl_215} shows the detailed SDSS
spectrum of the \lya, Ly$\beta$, and Lyman limit absorption spectrum
in pair 1, at the redshift of the foreground QSO. We see that despite
the broad \lya\ absorption line, there is no discernible Lyman limit
absorption, and Ly$\beta$ absorption only shows two narrow
subcomponents. This shows that the absorption does not arise from a
few highly dense clumps in a Lyman limit system, but from gas that is
well spread over the $2000 \kmVs$ velocity range. In fact, to have
saturated \lya\ absorption over this velocity range requires a minimum
column density of $\sim 10^{16} \cm^{-2}$, but the total column density
cannot be larger than $\sim 10^{17} \cm^{-2}$ due to the absence of
Lyman limit absorption. This suggests that the absorption system is due
to gas distributed over a region comparable in size to the impact
parameter to the QSO. Such a large region should have been well
resolved in our simulations.

  For pairs 2 and 3, the absorption is less strong and the factor
$1+\omega$ is smaller, so the possibility of a coincident Lyman limit
system is not as unlikely as in pair 1. We have also searched for
Lyman limit absorption in pair 2, and there is none, so the strong \lya\
absorption lines in pair 2 near the foreground QSO must have a total
column density below $\sim 10^{17}\cm^{-2}$ (for pair 3, the low
redshift of the foreground QSO puts the Lyman limit outside the
SDSS spectrum wavelength range).

  The fact that the transverse proximity effect is absent in all three
of our pairs essentially rules out that the absorption is caused by
random Lyman limit systems unrelated to the foreground QSOs, even if the
probabilities $P_<$ obtained from our simulations were too low by a
factor up to $10$ and if we ignore the additional difficulty of the
high velocity dispersion required. The evidence shows that either the
gas density is greatly enhanced out to distances of $\sim 1$ proper Mpc
from QSOs, or the QSO ionizing flux is suppressed because of variability
and/or anisotropy of the emission.

\section{Clustering}
\label{sect:clustering}

  We have assumed as a starting point that the average gas density in
the vicinity of the foreground QSO is not different from the overall
average density.  There is, however, reason to believe that there is a
density enhancement around both radio-loud and radio-quiet QSOs from
the galaxy-QSO clustering that is observed (see Ellingson, Green \& Yee
1991; Hutchings \etal 1995; McLure \& Dunlop 2001; Wold \etal 2001;
Finn \etal 2001) and has been modeled (Kauffman \& Haehnelt 2002).
This density enhancement would result in excess absorption, diminishing
the signature of the proximity effect due to the ionizing radiation,
which would lead to an \emph{over-estimate} of the background ionization
rate (see \figp \ref{fig:Gamma_eff}). Pascarelle \etal (2001) attempted
to use the observed QSO-galaxy correlation to correct the ionizing
background intensity inferred from the proximity effect, but their
conclusions depend on their assumption that all \lya\ lines originate in
fixed gaseous halos around galaxies, and they neglected peculiar
velocity effects.

  A gas density enhancement would also be expected around Lyman-Break
Galaxies (LBGs). Adelberger \etal (2003) found that the mean transmitted
flux decreases near an LBG, dropping from a mean value of $0.67$ at
large distances to $\sim0.55$ at $0.5h^{-1}$ proper Mpc, at $z=3$.
\footnote{Adelberger \etal also find high transmitted flux within
$0.125h^{-1}\Mpc$ (proper) from LBGs, which they attribute to galactic
wind effects, but this distance is much smaller than those probed by our
QSO pairs.}  Using our simulations, we find that this decrease in
mean transmitted flux corresponds to an increase in the optical depth by
a factor of $\sim2$. The optical depth in pair 1 is expected to decrease
from the enhanced photoionization rate by a factor of $(1+\omega)^{-1}
\sim1/90$ at this separation, so if the gas density enhancement around
LBGs and luminous QSOs were the same, it could not explain the lack of a
proximity effect in our data. Nevertheless, QSOs may generally inhabit
more massive halos than do LBGs, placing them in dense environments of
a larger scale. 

\begin{figure}[th!]
  \PSbox{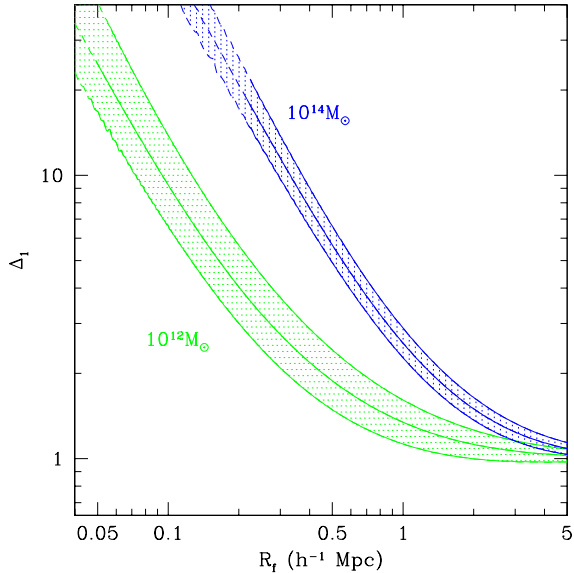  hoffset=70 voffset=-65 hscale=40 vscale = 40}
  {75mm}{75mm}
\caption{The non-linear overdensity $\Delta_1$ at a proper radius
$R_f$ around collapsed structures of mass $10^{12} \Msun$ and $10^{14}
\Msun$ at $z=4$. The shaded regions  bracket the one sigma deviation
in initial overdensity.  The solid lines turn to dashed lines when the
radius has passed below the virial radius, where we imagine the
overdensity will diverge from the spherical collapse model.
\label{fig:od_compare} }
\end{figure}

  Without an appropriate empirical estimate of the gas clustering
around QSOs, we will attempt to model it analytically, following the
calculations in Loeb \& Eisenstein (1995). As outlined in the
Appendix, we calculate (by assuming Gaussian fluctuations and the
spherical collapse model) the non-linear matter overdensity $\Delta_1$
at proper radius $R_f$ corresponding to the average density perturbation
around a halo (see \figp \ref{fig:od_compare}), and we assume that the
gas traces the matter, i.e., $\Delta_1=\rho_{gas}/\bar\rho_{gas}$. This
gas overdensity will
affect the neutral hydrogen density through the relation:  $\rho_{HI}
\propto \rho_{gas}^2 \alpha(T)$, where $\alpha(T) \propto T^{-0.7}$ is
the recombination coefficient. For $T\propto\rho_{gas}^{0.3}$, this
implies $\rho_{HI} \propto \rho_{gas}^{1.79}$, and including also the
effects of peculiar velocities, the optical depth will increase by:

\beq
  1+\omega_{cl}=\Delta_1^{1.79}
                \left|\left(1+\frac{d\v_{pec}}{Hd\ell}\right)^{-1}\right|,
\label{eq:clustering}
\eeq

\noindent
where $d\ell$ is the space line element. Note that caustics on
$\omega_{cl}$ appear when $d\v_{pec}/d\ell = -H$, and inside the caustics
we need to add the contribution to $1+\omega_{cl}$ from multiple points
in space to the same observed wavelength. The peculiar velocities shift
the wavelength at which we observe the absorption, thereby changing \eq 
(\ref{eq:R_ll}) to

\beq
  R_\parallel =  \frac{\Delta \v-\v_{pec}}{H(z_f)} ~.
\label{eq:R_ll_vpec}
\eeq 	

\noindent
Because $\v_{pec}$ has the opposite sign of
$R_\parallel$ (see Appendix), absorption observed at a
redshift separation $\Delta \v$ will be located at a distance
$R_f=(R_\bot^2+R_\parallel^2)^{1/2}$, which is in general farther than
in the absence of peculiar velocities.  The larger inferred distance
for a given $\Delta \v$ implies that the expected flux from the QSO,
as well as the clustering effect due to the enhanced gas density, are
reduced.

\begin{figure}[th!]
  \PSbox{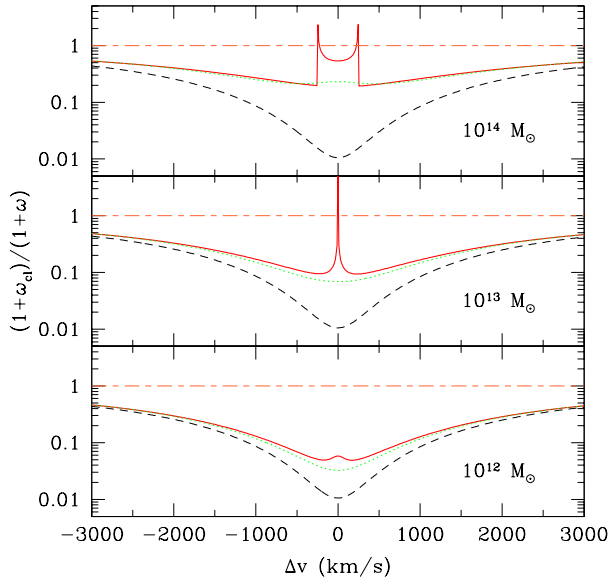  hoffset=70 voffset=-65 hscale=40 vscale = 40}
  {75mm}{75mm}
\caption{ {\it Solid lines:} Ratio of the clustering effect
(\eqp \ref{eq:clustering}) to the radiation effect (\figp \ref{fig:G_n_w})
in pair 1 for halo masses of  $10^{14} \Msun$ (top), $10^{13} 
\Msun$ (middle), and $10^{12} \Msun$ (bottom). {\it Dashed
line:} Photoionization effect only, shown as $(1+\omega)^{-1}$.
{ \it Dotted line:} Ratio without including the effect from the peculiar
velocities.
\label{fig:w_w} }
\end{figure}

  We show in \fig \ref{fig:w_w} as solid lines the ratio
$(1+\omega_{cl})/ (1+\omega)$, for pair 1, where $1+\omega$ is the
factor by which the optical depth is suppressed due to the increased
ionizing flux, and $1+\omega_{cl}$ is the factor by which the optical
depth is increased by the enhanced gas density, according to \eq
(\ref{eq:clustering}).  The three panels assume that the QSO is in the
center of a halo with mass of ($10^{14}$, $10^{13}$, $10^{12}$)
$\Msun$, with the mean density profile of \fig
\ref{fig:od_compare}. The dotted line shows the same ratio when
peculiar velocities are not included, and the dashed line shows $ 1/
(1+\omega)$, i.e., the effects of photoionization only, also with no
peculiar velocities. Note that for a halo mass $10^{13} \Msun$, the
line of sight to the background QSO in pair 1 just grazes the
turnaround radius of the halo around the foreground QSO, producing the
high central caustic.

  We see that the effects of clustering are important and they can
partly compensate for the increased ionizing flux. However, even for a
halo mass of $10^{14} \Msun$, the total factor by which the optical
depth is changed by the combined effects of photoionization and
clustering is still substantially less than unity, except in the narrow
region between the velocity caustics within $\sim 250 \kmVs$ of the QSO
redshift. As mentioned earlier, the peculiar velocities decrease both
$1+\omega_{cl}$ and $1+\omega$, but their ratio is only slightly
modified outside the velocity caustics.

  Plausible values of the mass of the halo around the QSO can be
constrained by the mass of the black hole required to power the QSO,
and by comparing the abundance of halos with the abundance of QSOs.
A lower limit to the black hole is obtained by assuming that the
luminosity is equal to the Eddington luminosity, in which case:

\beq 
  \frac{M_\bullet}{\Msun} \simeq 10^{-0.4M_B-1.6} 
\eeq

\noindent
where $\sim$ 10\% of the emission is assumed to be in the $B$ band
(see Elvis \etal 1994).  For the magnitude of the foreground QSO in
pair 1, $M_B=-26.9$,\footnote{We derive the absolute B magnitude from
the given $i^*$ magnitude, assuming $L_\nu\propto\nu^{-0.44}$ from the
composite spectrum in Vanden Berk \etal (2001).} we have
$M_\bullet\sim10^{9} \Msun$.  Similar mass estimates can be made for
the foreground QSO in pair 2 ($M_B=-26.7$) and pair 3 ($M_B=-26.4$).
If we assume the typical black hole to bulge mass ratio of 0.13\%
(Merritt \& Ferrarese 2001) and assume the bulge to halo mass ratio is
of order the baryon fraction ($\sim10\%$), then we obtain a halo mass,
$M_h$, of roughly $10^{13} \Msun$.

\begin{figure}[th!]
  \PSbox{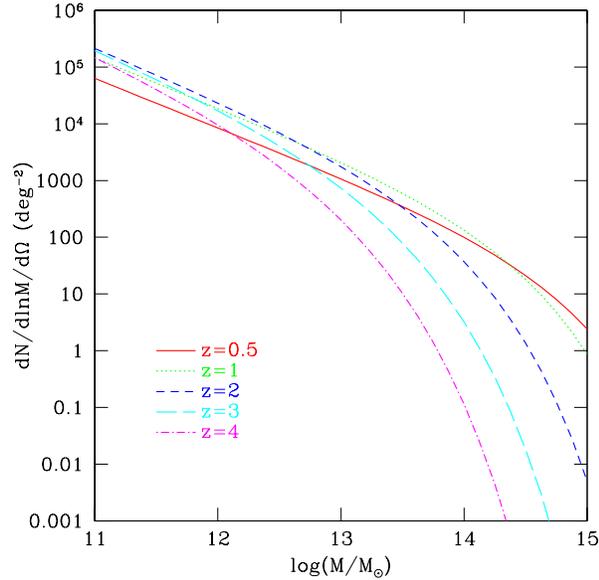  hoffset=70 voffset=-65 hscale=40 vscale = 40}
  {75mm}{75mm}
\caption{ The surface density per $\rm{deg}^2$ of halos of a given
mass per unit redshift from the Press-Schecter formalism.
Note that the cosmology assumed is the same as in the simulation:
$(\Omega_m, \Omega_\Lambda, h,\sigma_8, n_s)=(0.4,0.6,0.65,0.83,0.85)$.
\label{fig:all_halos} }
\end{figure}

  The surface density of halos per unit redshift has been plotted in \fig
\ref{fig:all_halos} using the Press-Schechter formalism (Press \&
Schechter 1974). At the redshift of pair 1 ($z=4$), there are
(1000, 10, 0.001) halos per ${\rm deg}^2$ with ($10^{12}$, $10^{13}$,
$10^{14}$) $\Msun$, respectively.  We can compare this to the
0.2 QSOs per ${\rm deg}^2$ in the Early Data Release at the same
redshift.\footnote{There are 482 QSOs with $z>2.2$ and 91 with
$3.5>z>4.5$ over a field of $494{\rm deg}^2$.}  This means that it is
highly unlikely that the foreground QSO in pair 1 may reside in a
$10^{14} \Msun$ halo. However, the overdensity around smaller
mass halos is apparently not sufficient to offset the purported flux
from the foreground QSO. As shown in \fig \ref{fig:w_w}, the effect
of clustering is too small to eliminate the effect of the extra
radiation in pair 1 except for the region between the caustics, with a
width of only $\Delta \v \sim 500\kmVs$ even for the case $M_h=10^{14}
\Msun$. For a $10^{13} \Msun$ halo (a more likely mass given the halo
abundances), the region between caustics has practically disappeared
and the optical depth should still be decreased by a factor of 5 to 10
over most of the region in which excess absorption is observed
($-1500\kmVs<\Delta \v< 500\kmVs$ in \figp \ref{fig:prob_tran_comp}).

\begin{figure}[th!]
  \PSbox{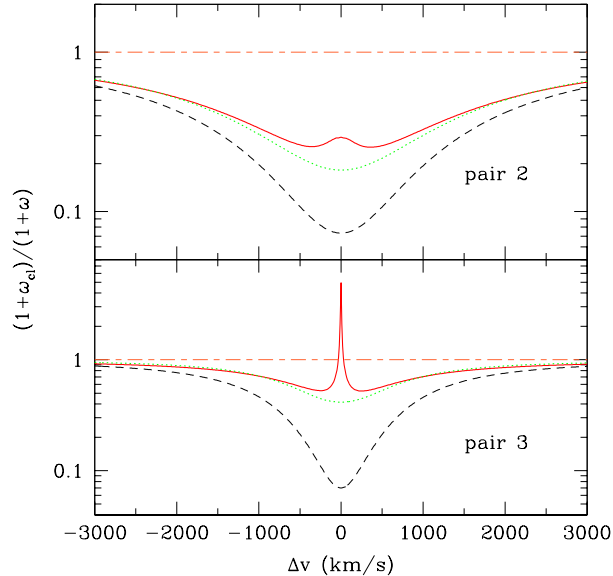  hoffset=70 voffset=-65 hscale=40 vscale = 40}
  {75mm}{75mm} 
\caption{ Ratio of the clustering effect to the
radiation effect as in \fig \ref{fig:w_w}, but here for pair 2
(\emph{top}) and pair 3 (\emph{bottom}), assuming for both a halo of
$10^{13} \Msun$.  
\label{fig:w_w_23} } 
\end{figure}

  For the other two pairs, we have plotted the clustering/radiation
ratio in \fig \ref{fig:w_w_23}, assuming a $10^{13} 
\Msun$ halo.  In the case of pair 2, the clustering again appears
to be insufficient for offsetting the ionizing flux from the
foreground QSO.  As for pair 3, it is possible that the reduced
absorption that we expected from the transverse proximity effect has
been mitigated by the enhanced density around the foreground QSO.
However, it might remain difficult to explain the wide absorption
feature seen in \fig \ref{fig:prob_tran_comp} at $\sim500\kmVs$ if
indeed the expected excess radiation is impinging on that region of
the IGM.  In any case, clustering by itself appears to be  unable to
account for the lack of a transverse proximity effect in all three
pairs.  The alternatives, which we address in the next sections, are
QSO variability and anisotropy.

\section{QSO anisotropy}

  A different possibility to explain the lack of a transverse proximity
effect is that the emission from QSOs is highly anisotropic. In any
flux limited survey, the observed QSOs would be preferentially observed
along directions in which they appear particularly bright, and their
flux seen along other directions could be much lower.

\begin{figure}[th!]
  \PSbox{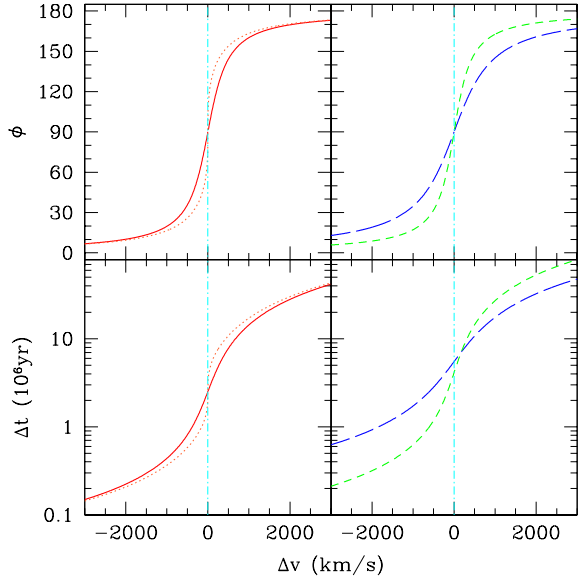  hoffset=70 voffset=-65 hscale=40 vscale = 40}
  {75mm}{75mm}  
\caption{ \emph{Top left}-- The angle, $\phi$, between the line of
sight to the foreground QSO and the line connecting the QSO to the
point at $\Delta\v$ in pair 1.  The \emph{solid} line assumes no
clustering, whereas the \emph{dotted} line assumes the foreground QSO
is in a halo of $10^{13} \Msun$.  We plot both cases to show the
effect of the induced peculiar velocities.  \emph{Top right} --  The
angle, $\phi$, for pair 2 (\emph{long dash}) and pair 3 (\emph{short
dash}) without clustering.  \emph{Bottom left and right}--  The delay
between the time when the QSO emits the light we see and the time when
the QSO emits the light impinging on the point at $\Delta\v$.  The
lines are the same as above.
\label{fig:phi_n_t} }  
\end{figure}

  The angle, $\phi$, that a given redshift separation corresponds to
is plotted in the top two panels of \fig\ref{fig:phi_n_t}.  
In the case of pair
1, the QSO flux would have to be obscured from roughly
$\phi\simeq10^\circ$ to $\phi\simeq150^\circ$ to explain the low
probability region between $\Delta\v=-1500 \kmVs$ and  $\Delta\v=500
\kmVs$.  If we assume that this obscuration is symmetric around an
axis, then the half-opening angle would only be $\sim 20^\circ$.  As
shown in \fig\ref{fig:phi_n_t}, peculiar velocities would have a
only a small effect on this angle.  For pairs 2 and 3, a wider
half-opening angle ($\sim 40^\circ-50^\circ$) is allowed because of
the larger impact parameter and the lower velocity range in the
spectrum in which the probability $P_<$ is very low.

  A beam radius as small as $20^\circ$ for the QSO radiation seems
implausible. In unified models of AGN, the continuum ionizing
radiation is supposed to come from the accretion disk, which may be
absorbed by an obscuring torus near the equator, but typical
half-opening angles are $\sim 30^\circ-45^\circ$ (Antonucci 1993;
Schmitt \etal 2001), and they are thought to increase with luminosity
(Rudge \& Raine 2000). A separate possibility is that the QSO has not
ionized the gas in its host halo, and that the ionizing radiation is
able to escape only along a narrow tunnel among clouds.  However, the
fact that most QSOs of luminosity similar to the foreground one in
pair 1 do not exhibit intrinsic Lyman limit absorption in their
spectrum implies that this explanation could not account for a narrow
beam of emission in most QSOs.

  To summarize, beaming of the ionizing radiation might be one of the
reasons for the absence of the transverse proximity effect in our three
pairs, but if this absence is generally confirmed on a larger sample of
pairs, then beaming alone cannot be the sole explanation.

\section{QSO variability}
\label{sect:variability}

  The final effect we consider is the possibility that the QSO flux is
variable. There is a time delay between the emission of the QSO flux
that we observe at present and the emission of the QSO flux that
illuminated the hydrogen in the spectrum of the background QSO, given
by $\Delta t \simeq \frac{1}{c}(R_f+R_\parallel)$, where $R_f$ and
$R_\parallel$ are defined in \eqs (\ref{eq:R_f}) and (\ref{eq:R_ll}).
The time-delay is plotted in the bottom two panels of \fig
\ref{fig:phi_n_t} for all three pairs.  The dotted curve shows the
effect of including peculiar velocities when the foreground QSO in
pair 1 is in a halo of mass $10^{13} \Msun$.

We again focus on the region $-1500\kmVs<\Delta \v< 500\kmVs$ in pair 1
where the probability $P_<$ is lowest.  If the lack of a proximity effect
were due to the time-delay only, we find by comparing \figs
\ref{fig:G_n_w} and \ref{fig:phi_n_t} that the QSO luminosity would
need to have been $\sim10$ times lower than now at a time $3\times
10^5$ years ago, and 50 to 100 times lower than now over the period
from $10^6$ to $10^7$ years ago.   In the case of pairs 2 and 3, the
QSOs would have to be shining an order of magnitude less bright from
$\sim2\times10^6$ to $10^7$ years ago.  These required variations in the
luminosities would be smaller if significant clustering existed around
the QSO (\figsp \ref{fig:w_w}  and \ref{fig:w_w_23}).  

  There are other measurements related to the proximity effect that
are sensitive to the time interval over which QSOs maintain their
luminosity. The existence of the line-of-sight proximity effect requires
that the QSOs shine continuously for a photoionization timescale, which
is $\sim 3\times 10^4$ years for the measured ionizing background
intensity (see the introduction). The case of Q0302-003, and its
\lya\ void associated with another nearby QSO, analyzed by Dobrzycki \&
Bechtold (1991), requires a lifetime longer than $10^7$ years if the
void is really caused by the QSO. The same QSO, Q0302-003, was observed 
in the \heii\ \lya\ spectrum by Jakobsen \etal (2003), who found a gap in
the absorption associated with another QSO near the line of sight.
The required QSO lifetime is also $10^7$ years if the QSO is responsible
for the gap. The large \hii\ region around a z=6.28 QSO also requires a
lifetime longer than $10^7$ years if it was entirely ionized by the QSO
(Pentericci \etal 2002; Haiman \& Cen 2002).

  We point out here another argument that shows that many luminous QSOs
must sustain their luminosities for periods of at least $10^7$ years.
Of the four QSOs on which the \heii\ \lya\ intergalactic absorption has
been observed, two show a strong line of sight proximity effect:
Q0302-003, and PKS1935 (Anderson \etal 1999, Heap \etal 2000). The
observed flux of these QSOs of $\sim 2\times 10^{-16}
{\rm erg}\,{\rm cm}^{-2}\,{\rm s}^{-1}\,\mbox{\AA}^{-1}$ implies a
photon luminosity of $\sim 10^{56}\, {\rm s}^{-1}$ in photons that can
ionize \heii\ (i.e., with energy greater than 54 eV). The size of the
proximity effect in the spectra of these two QSOs is $\Delta z\simeq
0.08$, or a proper distance of 20 Mpc. With this luminosity and at
this distance, the photoionization rate on \heii\ ions that is obtained
is $\sim 10^{-7} \,{\rm yr}^{-1}$. If the QSO had been shining for a
time less than $10^7$ years, it could not have ionized the region
in which the proximity effect is observed to the level implied by its
luminosity (see Anderson \etal 1999). We note that this argument does
not require the assumption that the QSO {\it ionized} the entire \heiii\
region, but only that the QSO has maintained its ionization over the
last photoionization timescale.

  It is worth mentioning here that one must distinguish between the
total length of time during which a black hole is producing a certain
luminosity as it accretes, and the continuous time interval over which
a QSO maintains its luminosity. QSOs might have repeated episodes of
high luminosity, each one relatively short and separated by intervals
of lower luminosity, which could add up to a longer total time of
emission. Several papers (Fabian \& Iwasawa 1999; Barger \etal 2001;
Yu \& Tremaine 2002; Haehnelt 2003) have shown that the total mass
density of nuclear black holes in the universe is comparable to the
mass density inferred from the radiation background from QSOs (Soltan
1982), for an assumed radiative efficiency of the accreted matter:
$\epsilon\sim10\%$.  This also implies that, if QSOs generally emit
near the Eddington luminosity, $L_{Edd}$, then their total time of
emission is $L_{Edd}/(M_\bullet c^2\epsilon)\simeq 4\times 10^7$
years. Measurements based on the amplitude of QSO clustering and the
assumption that QSOs are located in massive halos (Martini \& Weinberg
2001; Martini 2003) also measure this total time of emission. 

  In conclusion, we have found that although the time-delay effect can
explain the absence of the proximity effect, it requires the
foreground QSO in pair 1 to be younger than $10^6$ years, and the
lifetime should be $< 2\times 10^6$ years for pairs 2 and 3 (owing to
their larger impact parameters). At the same time, there is evidence
that other QSOs have sustained their luminosities for much longer
times. Therefore, time-delays can only eliminate the transverse
proximity effect in some QSOs if there is a wide range of variation in
the time intervals over which QSOs remain at a high luminosity.

\section{Conclusion}

  We have found three sets of SDSS QSO pairs in which we expected to see
evidence for the transverse proximity effect. The fact that we do not
implies that either the gas density along the line of sight is much
higher than average, or the ionizing flux expected from the observed QSO
is suppressed.  

  Although it is likely that the high-redshift, luminous QSOs are
located near the center of the most massive halos that exist at the time
when they are observed, we have found that the gas density enhancement
is not sufficient to fully compensate for the increased ionizing
intensity from the QSO. For reasonable halo masses that are consistent
with the abundance of QSOs, the impact parameters in our QSO pairs are
similar or larger than the radius around the halo where gas is just
turning around, and the overdensity is not large. This suggests that
there must also be a reduction of the QSO flux, either due to
anisotropic emission or to a lower luminosity of the QSO in the past
$3\times 10^5$ to $10^7$ years. Each one of these two explanations
also has problems to account for the absence of a transverse proximity
effect in three pairs by itself: anisotropy requires a very small
opening angle, and variability implies a large reduction of the
luminosity over a timescale over which other QSOs are known to have
remained highly luminous (Jakobsen \etal 2003). The
combination of the three effects (high gas density around the QSOs,
anisotropy and variability) can probably more easily account for the
observed \lya\ excess absorption near the foreground QSOs, rather than
the excess transmission that was expected.

  Previous studies have also generally found no evidence for the
transverse proximity effect.  Crotts \& Fang (1998) found, just as we do,
excess \lya\ absorption compared to the mean transmitted flux in the
region of the proximity effect.  M\o ller \& Kj\ae rgaard (1992) concluded
from an analysis of one pair and the triplet of QSOs from Crotts (1989)
that the transverse flux from local QSOs had to be reduced by a factor of at
least $0.22$ for an assumed background photoionization rate
$\Gamma_{bkg}\simeq 3\times 10^{-12} \sec^{-1}$. For our lower assumed
values of $\Gamma_{bkg}$ (\figp \ref{fig:Gamma_eff}), this corresponds to a reduction
factor $\sim0.05$. Fern\'andez-Soto \etal (1995) detected in 3 pairs a
weak transverse proximity effect that was consistent with a high value of
the background ionization rate, $\Gamma_{bkg}\simeq 10^{-11} \sec^{-1}$.
Assuming this rate is an order of magnitude too large,
we can interpret the small observed effect as due to
a reduction in the transverse flux by a factor of $0.1$.
To express our results in terms of a reduction factor, we can assume that
the probability $P_<$ for all points in
\fig\ref{fig:prob_tran_comp} should, minimally,
be greater than 0.001.  To obtain this, we find the transverse flux
in pair 1 must be less than 0.07 times the expected isotropic value.
The fact that many of the closest pairs in the literature exhibit at least
a 10\% reduction in the transverse flux considerably strengthens
our conclusion.

  Future progress in understanding the proximity effect should 
come from a reanalysis of the
line-of-sight proximity effect in a large sample of QSOs. An enhanced
gas density near QSOs should equally impact the proximity effect on the
line of sight, and although its strength is complicated by the effect of
peculiar velocities this can be modeled in simulations. The environment
of QSOs can be probed not only from the average proximity effect, but
also from its scatter. It is well known that the value of the ionizing
background intensity inferred from the proximity effect varies widely
from one QSO to another, implying that a large sample must be used;
however, this scatter is probably not explained by the \lya\ absorption
variance in random regions of the intergalactic medium, and the effect
of the QSO environment is likely to be important.

  A thorough investigation of the line-of-sight proximity effect should
help in disentangling the effects of gas density and the reduction of
the QSO flux in the transverse proximity effect. The effects of
anisotropy and variability can hopefully also be separated, once a large
number of pairs over a range of angular separations are available. The
flux reduction due to anisotropy should depend only on the angle $\phi$
(see \S 6) and not the distance to the QSO, and should be symmetric when
changing $\phi$ to $\pi-\phi$ (except if the two QSO ``beams'' are often
of different luminosity), whereas the mean reduction due to variability
should be a monotonically increasing function of time delay. There is a
promising potential in the transverse proximity effect for learning on
the environment, anisotropy, and variability of QSOs.

\emph{Acknowledgements.}  We thank Gordon Richards for
supplying us with the list of high redshift SDSS QSOs.  We also thank
James Bullock, Juna Kollmeier, Smita Mathur, Pat Osmer, Rick Pogge,
Terry Walker, David Weinberg, and Andrew Zentner for helpful
discussions. We acknowledge support from U.S. DOE Contract
No.DE-FG02-91ER40690 for MS, and from NSF grant NSF-0098515 for JM.
The numerical simulations used in this paper were
performed at NCSA.

\section{Appendix}

To determine the overdensity as a function of radius outside a
collapsed object, we assume the spherical collapse model to determine
the evolution of the proper radius $R$ of a shell containing a
constant mass $M$, which is parameterized as:  

\beq
  R = \frac{R_{ta}}{2} (1-\cos(\eta)), \qquad  
  t = \frac{t_{ta}}{\pi}(\eta-\sin(\eta))
\label{eq:parm_Rnt}
\eeq

\noindent
where ``$ta$'' stands for \emph{turnaround}, i.e., the point of
maximum expansion for a particular shell.  We assume that the
foreground QSO is at the center of a halo with mass $M_h$ collapsed at
the time we are viewing the QSO, at redshift $z_f$. In the standard
top-hat spherical model, this implies that the mean linear overdensity
inside the comoving radius $r_h\equiv[3M_h/(4\pi
\bar\rho_{m,0})]^{1/3}$ (where $\bar\rho_{m,0}$ is the mean comoving
matter density) is $\bar{\delta}_h=1.69$ (assuming $\Omega_m=1$, valid
at high redshift). To evaluate the average overdensity around the
halo, we follow the method of Loeb \& Eisenstein (1995). At a comoving
radius $r_1>r_f$, the normalized average linear density fluctuation is
$\nu_1\equiv\bar\delta_1/\sigma_M(r_1,z_f)$, where $\sigma_M(r_1,z_f)$
is the rms fluctuation $\delta M / M$ on a sphere of radius $r_1$, and
$\bar\delta_1$ is the mean linear overdensity within radius $r_1$
around the halo of mass $M_h$. The mean and the variance of the
distribution of $\nu_1$ are $\gamma_{h1}\cdot\nu_h$ and
$1-\gamma_{h1}^2$, respectively, where
$\nu_h\equiv\bar\delta_h/\sigma_M(r_h,z_f)$ and

\beq
  \gamma_{h1} = \frac{1}{2\pi^2\sigma_M(r_h)\sigma_M(r_1)}
                \int^\infty_0 k^2dk P(k) W(k \cdot r_h) W(k \cdot r_1) ~.
\eeq

\noindent
Here, $W(x) = 3j_1(x)/x$ is the Fourier window function for
a top-hat filter, and $j_1$ is a spherical Bessel function.
\footnote{Eq.\ 8 in Loeb \& Eisenstein (1995) appears to be missing a
factor of $3/(\sqrt{2}\pi)$.  This assures that $\gamma_{h1}=1$ for
$r_h=r_1$.} 

  We next solve for the non-linear overdensity
$\Delta_1\equiv\rho_m(R_1)/\bar\rho_m$, where $\rho_m(R_1)$ is the
proper matter density at the radius $R_1$ implied by the linear
overdensity $\bar\delta_1$, and $\bar\rho_m=\bar\rho_{m,0}
(1+z_f)^3$. The turnaround radius and turnaround time of the shell at
$r_f$ are (in a flat cosmology)

\beq 
  R_{ta,1} = R_{ta,h} \frac{r_1 \bar{\delta}_h}{r_h \bar{\delta}_1} ~,
  \qquad \qquad
  t_{ta,1} = t_{ta,h} 
  \left(\frac{\bar{\delta}_1}{\bar{\delta}_h}\right)^{-3/2} ~.
\label{eq:ta_ratios} 
\eeq

\noindent
We solve for $\eta_1$ in equations (\ref{eq:parm_Rnt}) by
equating the time variable for the shells at $r_h$ and $r_1$, and then
find the radius $R_1$ of the shell in the non-linear evolution.  Note
that $R_1$ is the same as $R_f$ from \eq (\ref{eq:R_f}).  
The non-linear overdensity is

\beq
  \Delta_1 = \left(\frac{r_1}{R_1}\right)^{2}
                  \frac{dr_1}{dR_1} \ (1+z_f)^{-3} ~.
\eeq

\noindent 
We have plotted this overdensity in \fig\ref{fig:od_compare} 
for two mass scales, $10^{12}{\rm M_\odot}$ and
$10^{14}{\rm M_\odot}$, and we assume in \eq (\ref{eq:clustering}) that
the gas overdensity is equal to the matter overdensity.  The velocity of 
the shell in the reference frame of the center of the  sphere is:

\beq
  \v_1 \equiv  \frac{dR}{dt} =
            \pm H \ r_1 \ \sqrt{1-\frac{R_1}{R_{ta,1}}} \ 
                \left(\frac{r_1}{R_1}\right)^{1/2} \ (1+z_f)^{-3/2} 
\eeq

\noindent
where by definition we make $\v_1$ negative when a shell
is collapsing (i.e. after turnaround).  The peculiar velocity along
the line of sight to the background QSO will be
$\v_{pec}=-R_\parallel\cdot(H - \v_1/R_1)$, where
$R_\parallel=(R_1^2-R_\bot^2)^{1/2}$ and $R_\bot$ is the impact
parameter.

\end{document}